\documentclass[showpacs,preprintnumbers,amsmath,amssymb,aps,prl,preprint]{revtex4-1}

\usepackage{graphicx}
\usepackage{dcolumn}
\usepackage{bm}
\usepackage{color}
\usepackage{times}
\usepackage{wasysym}
\renewcommand{\vec}[1]{\mathbf{#1}}
\usepackage{graphicx}
\usepackage{times}
\usepackage{multirow}

\newcommand{\vE}{\ensuremath{\mathbf{E}}}
\renewcommand{\vr}{\ensuremath{\mathbf{r}}}

\newcommand{\w}{\ensuremath{\omega}}

\newcommand{\lp}{\ensuremath{\left(}}
\newcommand{\rp}{\ensuremath{\right)}}
\renewcommand{\a}{\ensuremath{\alpha}}

\newcommand{\chieff}{\ensuremath{\chi^{(2)}_\text{eff}}}

\newcommand{\Qc}{\ensuremath{Q^\text{c}}}

\renewcommand{\eqref}[1]{(\ref{eq:#1})}
\newcommand{\eqreftwo}[2]{(\ref{eq:#1}), (\ref{eq:#2})}

\newcommand{\figref}[1]{Fig.~\ref{fig:#1}}
\newcommand{\Figref}[1]{Figure~\ref{fig:#1}}

\begin{document}

\title{Cavity-enhanced second harmonic generation via nonlinear-overlap optimization}

\author{Zin Lin$^1$}
\email{zlin@seas.harvard.edu}
\author{Xiangdong Liang$^2$}
\author{Marko Loncar$^1$}
\author{Steven G. Johnson$^2$}
\author{Alejandro W. Rodriguez$^{3}$}%
\affiliation{$^1$School of Engineering and Applied Sciences, Harvard University, Cambridge, MA 02138}
\affiliation{$^2$Department of Mathematics, Massachusetts Institute of Technology, Cambridge, MA 02139}
\affiliation{$^3$Department of Electrical Engineering, Princeton University, Princeton, NJ, 08544}

\date{\today}

\begin{abstract}
  We describe an approach based on topology optimization that enables
  automatic discovery of wavelength-scale photonic structures for
  achieving high-efficiency second-harmonic generation (SHG). A key
  distinction from previous formulation and designs that seek to
  maximize Purcell factors at individual frequencies is that our
  method not only aims to achieve frequency matching (across an entire
  octave) and large radiative lifetimes, but also optimizes the
  equally important nonlinear--coupling figure of merit $\bar{\beta}$,
  involving a complicated spatial overlap-integral between modes. We
  apply this method to the particular problem of optimizing micropost
  and grating-slab cavities (one-dimensional multilayered structures)
  and demonstrate that a variety of material platforms can support
  modes with the requisite frequencies, large lifetimes $Q >
  10^4$, small modal volumes $\sim (\lambda/n)^3$, and extremely large
  $\bar{\beta} \gtrsim 10^{-2}$, leading to orders of magnitude
  enhancements in SHG efficiency compared to state of the art photonic
  designs. Such giant $\bar{\beta}$ alleviate the need for
  ultra-narrow linewidths and thus pave the way for wavelength-scale
  SHG devices with faster operating timescales and higher tolerance to
  fabrication imperfections.
\end{abstract}


\pacs{Valid PACS appear here}%
\maketitle

{\it Introduction.---} Nonlinear optical processes mediated by
second-order ($\chi^{(2)}$) nonlinearities play a crucial role in many
photonic applications, including ultra-short pulse
shaping~\cite{DeLong94, Arbore97}, spectroscopy~\cite{Heinz82},
generation of novel frequencies and states of
light~\cite{Kuo06,Vodopyanov06,Krischek10}, and quantum information
processing~\cite{Vaziri02, Tanzilli05, Zaske12}.  Because
nonlinearities are generally weak in bulk media, a well-known approach
for lowering the power requirements of devices is to enhance nonlinear
interactions by employing optical resonators that confine light for
long times (dimensionless lifetimes $Q$) in small volumes
$V$~\cite{JoannopoulosJo08-book,Soljacic02:bistable,Soljacic03:OL,Yanik03,Yanik04,Bravo-AbadRo10,Rivoire09,Pernice12,Bi12,Buckley14}. Microcavity
resonators designed for on-chip, infrared applications offer some of
the smallest confinement factors available, but their implementation
in practical devices has been largely hampered by the difficult task
of identifying wavelength-scale ($V \sim \lambda^3$) structures
supporting long-lived, resonant modes at widely separated wavelengths
and satisfying rigid frequency-matching and mode-overlap
constraints~\cite{Rodriguez07:OE,Bravo-AbadRo10}.

In this letter, we extend a recently proposed formulation for the
scalable topology optimization of microcavities, where every pixel of
the geometry is a degree of freedom, to the problem of designing
wavelength-scale photonic structures for second harmonic generation
(SHG). We apply this approach to obtain novel micropost and grating
microcavity designs supporting strongly coupled fundamental and
harmonic modes at infrared and visible wavelengths with relatively
large lifetimes $Q_{1},Q_2 > 10^4$. In contrast to recently
proposed designs based on known, linear cavity structures
hand-tailored to maximize the Purcell factors or mode volumes of
individual resonances, e.g. ring
resonators~\cite{Almeida04,Xu05,Levy11,Pernice12} and nanobeam
cavities~\cite{Deotare09,Buckley14}, our designs ensure frequency
matching and small confinement factors while also simultaneously
maximizing the SHG enhancement factor $Q_1^2 Q_2 |\bar{\beta}|^2$ to
yield orders of magnitude improvements in the nonlinear coupling
$\bar{\beta}$ described by \eqref{beta} and determined by a special
overlap integral between the modes. These particular optimizations of
multilayer stacks illustrate the benefits of our formalism in an
approachable and experimentally feasible setting, laying the framework
for future topology optimization of 2D/3D slab structures that are
sure to yield even further improvements.

\begin{table*}[ht!]
\centering
\resizebox{\textwidth}{!}{
\begin{tabular}{l|c|c|c|c|c|c|c}
  \hline
  Structure                    & $h_x \times h_y \times h_z~(\lambda_1^3)$ & $\lambda~(\mathrm{\mu m})$ & ($Q_1,~Q_2$) & ($Q_1^\text{rad},~Q_2^\text{rad}$) 	& $\bar{\beta}$ & FOM$_1$           	& FOM$_2$           \\ \hline
  (1) AlGaAs/Al$_2$O$_3$ micropost & $8.4 \times 3.5 \times 0.84$       & 1.5 -- 0.75                & (5000, 1000) & ($1.4 \times 10^5$, $1.3 \times 10^5$) & 0.018         & $7.5 \times 10^6$     & $8.3 \times 10^{11}$  \\
  (2) GaAs gratings in SiO$_2$     & $5.4 \times 3.5 \times 0.60$       & 1.8 -- 0.9                 & (5000, 1000) & ($5.2 \times 10^4$, 7100)          	& 0.020         & $7 \times 10^6$   	& $7.5 \times 10^9$ \\
  (3) LN gratings in air           & $5.4 \times 3.5 \times 0.80$       & 0.8 -- 0.4                 & (5000, 1000) & (6700, 2400)                       	& 0.030         & $8.4 \times 10^5$ 	& $9.7 \times 10^7$ \\ \hline
\end{tabular}
}
\caption{SHG figures of merit for topology-optimized micropost and grating cavities of different material systems.}
\label{tab2}
\end{table*}

Most experimental demonstrations of SHG in chip-based photonic
systems~\cite{Rivoire09, Furst10, Levy11, Pernice12, Diziain13,
  Wang14, Kuo14} operate in the so-called small-signal regime of weak
nonlinearities, where the lack of pump depletion leads to the
well-known quadratic scaling of harmonic output power with incident
power~\cite{Boyd92}. In situations involving all-resonant conversion,
where confinement and long interaction times lead to strong
nonlinearities and non-negligible down
conversion~\cite{Soljacic03:OL,Rodriguez07:OE}, the maximum achievable
conversion efficiency $\left( \eta \equiv {P_2^\text{out} \over P_1^\text{in}} \right)$,
\begin{equation}
  \eta^\text{max}~= \left(1~-~\frac{Q_1}{Q_1^\text{rad}} \right) \left(1~-~\frac{Q_2}{Q_2^\text{rad}}\right)
\end{equation}
occurs at a critical input power~\cite{Rodriguez07:OE},
\begin{equation}
P_1^\text{crit} = {2 \w_1 \epsilon_0 \lambda_1^3 \over
  \big(\chieff\big) ^{2} |\bar{\beta}|^2 Q_1^2 Q_2}
\left(1-\frac{Q_1}{Q_1^\text{rad}} \right)^{-1},
\end{equation}
where $\chieff$ is the effective nonlinear susceptibility of the
medium [SM], $Q = \left(\frac{1}{Q^\text{rad}} +
\frac{1}{\Qc}\right)^{-1}$ is the dimensionless quality factor
(ignoring material absorption) incorporating radiative decay
$\frac{1}{Q^\text{rad}}$ and coupling to an input/output channel
$\frac{1}{\Qc}$. The dimensionless coupling coefficient $\bar{\beta}$
is given by a complicated, spatial overlap-integral involving the
fundamental and harmonic modes [SM],
\begin{align}
\bar{\beta} = { \int d\vr ~ \bar{\epsilon}(\vr) E_2^* E_1^2 \over
  \left( \int d\vr~ \epsilon_1 |\mathbf{E}_1|^2 \right) \left(
  \sqrt{\int d\vr~ \epsilon_2 |\mathbf{E}_2|^2 } \right) } \sqrt{
  \lambda_1^3 },
\label{eq:beta}
\end{align}
where $\bar{\epsilon}(\vr) = 1$ inside the nonlinear medium and zero
elsewhere. Based on the above expressions one can define the
following dimensionless figures of merit,
\begin{align} 
\label{eq:F1}
  \mathrm{FOM}_1 &= Q_1^2 Q_2 |\bar{\beta}|^2 \left( 1 - {Q_1 \over
    Q_1^\text{rad}} \right)^2 \left( 1 - {Q_2 \over Q_2^\text{rad}}
  \right), \\ \mathrm{FOM}_2 &= \left(Q^\text{rad}_1\right)^2
  Q^\text{rad}_2 |\bar{\beta}|^2
\label{eq:F2}.  
\end{align}
where $\mathrm{FOM}_1$ represents the efficiency per power, often quoted
in the so-called undepleted regime of low-power conversion~\cite{Boyd92}, and
$\mathrm{FOM}_2$ represents limits to power enhancement. Note
that for a given radiative loss rate, $\mathrm{FOM}_1$ is maximized
when the modes are critically coupled, $Q=\frac{Q^\text{rad}}{2}$,
with the absolute maximum occurring in the absence of radiative
losses, $Q^\text{rad} \to \infty$, or equivalently, when $\mathrm{FOM}_2$
is maximized. From either FOM, it is clear that apart from frequency
matching and lifetime engineering, the design of optimal SHG cavities
rests on achieving a large nonlinear coupling $\bar{\beta}$.

\begin{figure}[t!]
\centering
\includegraphics[width=0.5\columnwidth]{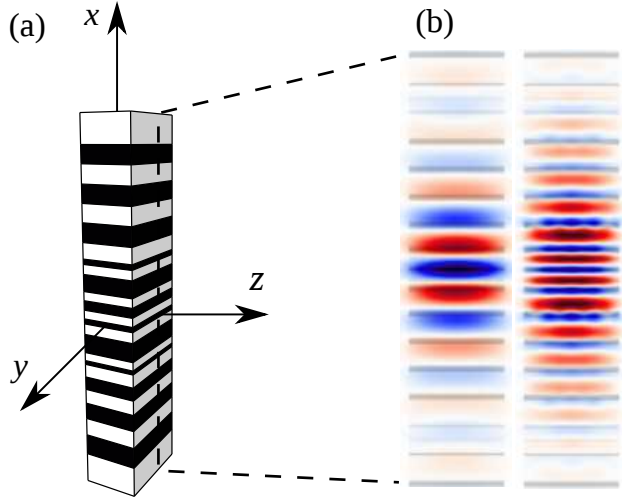}
\caption{(a) Schematic illustration of a topology-optimized micropost
  cavity with alternating AlGaAs/Al$_2$O$_3$ layers and dimensions
  $h_x \times h_y \times h_z = 8.4 \times 3.5 \times
  0.84~(\lambda_1^3)$. For detailed structural specifications, please
  refer to the supplement.  (b) $x$--$z$ cross-section of the $E_y$
  components of two localized modes of frequencies $\lambda_1 =
  1.5\mu$m and $\lambda_2 = \lambda_1/2$ and large spatial overlap
  $\sim E_2^* E^2_1$.}
\label{fig:fig1}
\end{figure}

\begin{figure}[ht!]
\centering
\includegraphics[width=0.5\columnwidth]{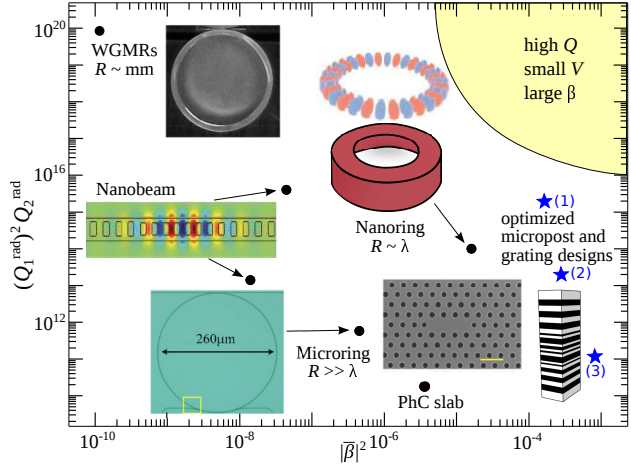}
\caption{Scatter plot of $ \left( Q_1^\text{rad} \right)^2 Q_2^\text{rad}$
  versus nonlinear overlap $|\bar{\beta}|^2$ for representative
  geometries, including WGMRs~\cite{Furst10}, micro- and nano-ring resonators~\cite{Pernice12,Bi12}, photonic crystal slab and nanobeam cavities~\cite{Rivoire09,Buckley14}. A trend towards decreasing lifetimes and increasing
  overlaps is readily observed as devices become increasingly smaller. 
  Meanwhile, it remains an
  open problem to discover structures with high $Q$s, small $V$s and
  large $|\bar{\beta}|$ (shaded region).
  \label{fig:fig2}}
\end{figure}

{\it Optimal designs.---} Table I characterizes the FOMs of some of
our newly discovered microcavity designs, involving simple micropost
and gratings structures of various $\chi^{(2)}$ materials, including
GaAs, AlGaAs and LiNbO$_3$. The low-index material layers of the
microposts consist of alumina (Al$_2$O$_3$), while gratings are
embedded in either silica or air (see supplement for detailed
specifications). Note that in addition to their performance
characteristics, these structures are also significantly different
from those obtained by conventional methods in that traditional
designs often involve rings~\cite{Pernice12, Bi12}, periodic structures
or tapered defects~\cite{Deotare09}, which tend to ignore or sacrifice
$\bar{\beta}$ in favor of increased lifetimes and for which it is also
difficult to obtain widely separated
modes~\cite{Buckley14}. \Figref{fig1} illustrates one of the optimized
structures---a doubly-resonant rectangular micropost cavity with
alternating AlGaAs/Al$_2$O$_3$ layers---along with spatial profiles of
the fundamental and harmonic modes. It differs from conventional
microposts in that it does not consist of periodic bi-layers yet it
supports two localized modes at precisely $\lambda_1= 1.5~\mu$m and
$\lambda_2=\lambda_1/2$. In addition to having large $Q^\text{rad} \gtrsim 10^5$
and small $V \sim (\lambda_1/n)^3$, the structure exhibits an
ultra-large nonlinear coupling $\bar{\beta} \approx 0.018$ that is
almost an order of magnitude larger than the best overlap found in the
literature (see~\figref{fig2}). From an experimental point of view,
the micropost system is of particular interest because it can be
realized by a combination of existing fabrication techniques such as
molecular beam epitaxy, atomic layer deposition, selective oxidation
and electron-beam lithography~\cite{Vahala04}. Additionally, the
micropost cavity can be naturally integrated with quantum dots and
quantum wells for cavity QED applications~\cite{Lermer12}. Similar to
other wavelength-scale structures, the operational bandwidths of these
structures are limited by radiative losses in the lateral
direction~\cite{JoannopoulosJo08-book,Vahala04,YZhang09}, but their
ultra-large overlap factors more than compensate for the increased
bandwidth, which ultimately may prove beneficial in experiments
subject to fabrication imperfections and for large-bandwidth
applications~~\cite{DeLong94, Arbore97, Scalora97, Krischek10}.

To understand the mechanism of improvement in $\bar{\beta}$, it is
instructive to consider the spatial profiles of interacting
modes. \Figref{fig1}b plots the $y$-components of the electric fields
in the $xz$-plane against the background structure. Since
$\bar{\beta}$ is a \emph{net} total of positive and negative
contributions coming from the local overlap factor $E_1^2 E_2$ in the
presence of nonlinearity, not all local contributions are useful for
SHG conversion. Most notably, one observes that the positions of
negative anti-nodes of $E_2$ (light red regions) coincide with either
the nodes of $E_1$ or alumina layers (where $\chi^{(2)}=0$),
minimizing negative contributions to the integrated overlap. In other
words, improvements in $\bar{\beta}$ do not arise purely due to tight
modal confinement but also from the constructive overlap of the modes
enabled by the strategic positioning of field extrema along the
structure.

Based on the tabulated FOMs (Table I), the efficiencies and power
requirements of realistic devices can be directly calculated. For
example, assuming $\chi^{(2)}_\text{eff}\left(\text{AlGaAs}\right) \sim
100~\mathrm{pm/V}$~\cite{Bi12}, the AlGaAs/Al$_2$O$_3$ micropost cavity~(\figref{fig1})
 yields an efficiency of
${P_{2,\mathrm{out}} \over P_1^2}=2.7\times 10^4/\mathrm{W}$ in the undepleted
regime when the modes are critically coupled,
$Q=\frac{Q^\text{rad}}{2}$.  For larger operational bandwidths,
e.g. $Q_1=5000$ and $Q_2=1000$, we find that ${P_{2,\mathrm{out}}
  \over P_1^2}=16/\mathrm{W}$. When the system is in the depleted
regime and critically coupled, we find that a maximum efficiency of
25\% can be achieved at $P^\mathrm{crit}_1 \approx 0.15~\mathrm{mW}$
whereas assuming smaller $Q_1=5000$ and $Q_2=1000$, a maximum
efficiency of $96\%$ can be achieved at $P^\mathrm{crit}_1 \approx
0.96~\mathrm{W}$.

{\it Comparison against previous designs.---} Table II summarizes
various performance characteristics, including the aforementioned FOM,
for a handful of previously studied geometries with length-scales
spanning from $\mathrm{mm}$ to a few wavelengths (microns). Fig 2
demonstrates a trend among these geometries towards increasing
$\bar{\beta}$ and decreasing $Q^\text{rad}$ as device sizes decrease.  Maximizing
$\bar{\beta}$ in millimeter-to-centimeter scale bulky media translates
to the well-known problem of phase-matching the momenta or propagation
constants of the modes~\cite{Boyd92}. In this category, traditional
WGMRs offer a viable platform for achieving high-efficiency
conversion~\cite{Furst10}; however, their ultra-large lifetimes
(critically dependent upon material-specific polishing techniques),
large sizes (millimeter length-scales), and extremely weak nonlinear
coupling (large mode volumes) render them far-from optimal chip-scale
devices. Although miniature WGMRs such as microdisk and microring
resonators~\cite{Pernice12, Diziain13, Kuo14} show increased promise
due to their smaller mode volumes, improvements in $\bar{\beta}$ are
still hardly sufficient for achieving high efficiencies at low
powers. Ultra-compact nanophotonic resonators such as the recently
proposed nanorings~\cite{Bi12}, 2D photonic crystal
defects~\cite{Rivoire09}, and nanobeam cavities~\cite{Buckley14},
possess even smaller mode volumes but prove challenging for design due
to the difficulty of finding well-confined modes at both the
fundamental and second harmonic frequencies~\cite{Rivoire09}. Even
when two such resonances can be found by fine-tuning a \emph{limited}
set of geometric parameters~\cite{Bi12, Buckley14}, the
frequency-matching constraint invariably leads to sub-optimal spatial
overlaps which severely limits the maximal achievable $\bar{\beta}$.

\begin{figure}[ht!]
\centering
\includegraphics[width=0.5\columnwidth]{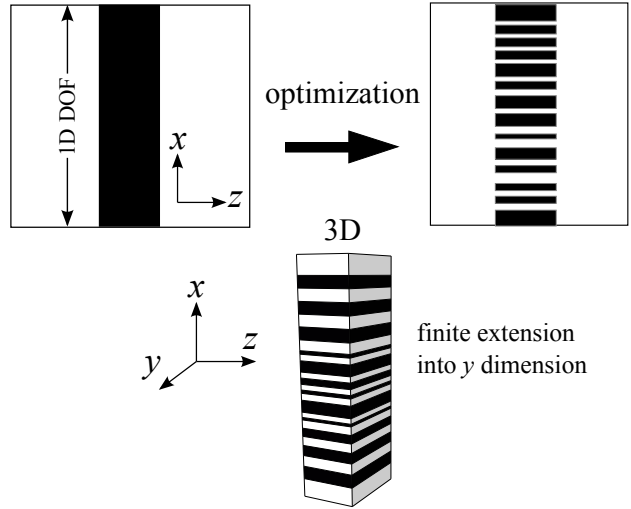}
\caption{Work flow of the design process. The degrees of freedom in
  our problem consist of all the pixels along $x$-direction in a 2D
  computational domain. Starting from the
  vacuum or a uniform slab, the optimization seeks to develop an
  optimal pattern of material layers (with a fixed thickness in the $z$-direction)
  that can tightly confine light at the desired frequencies while
  ensuring maximal spatial overlap between the confined modes. The
  developed 2D cross-sectional patterns is truncated at a finite width
  in the $y$-direction to produce a fully three-dimensional micropost or grating
  cavity which is then simulated by FDTD methods to extract the 
  resonant frequencies, quality factors, eigenmodes and corresponding
  modal overlaps. Here, it must be emphasized that we merely performed
  one-dimensional optimization (within a 2D computational problem)
  because of limited computational resources; consequently, our design space
  is severely constrained.~\label{fig:fig3}
  }
\end{figure}

Comparing Tables I and II, one observes that for a comparable $Q$, the
topology-optimized structures perform significantly better in both
$\mathrm{FOM}_1$ and $\mathrm{FOM}_2$ than any conventional geometry,
with the exception of the LN gratings, whose low $Q^\mathrm{rad}$ lead
to slightly lower $\mathrm{FOM}_2$. Generally, the optimized microposts and gratings perform better by
virtue of a large and robust $\bar{\beta}$ which, notably, is
significantly larger than that of existing designs. Here, we have not
included in our comparison those structures which achieve
non-negligible SHG by special poling techniques and/or quasi-phase
matching methods~\cite{Boyd92, Miller97, Kuo14}, though their
performance is still sub-optimal compared to the topology-optimized
designs. Such methods are highly material-dependent and are thus not
readily applicable to other material platforms; instead, ours is a
purely geometrical \emph{topology optimization} technique applicable
to any material system.


\begin{table*}[ht!]
\centering
\begin{tabular}{|c|c|c|l|c|l|c|c|c|}
  \hline 
  Structure    		& $\lambda~(\mathrm{\mu m})$ & \multicolumn{2}{c|}{($Q_1,~Q_2$)}             & \multicolumn{2}{c|}{($Q_1^\text{rad},~Q_2^\text{rad}$)} & $\bar{\beta}$ & $\mathrm{FOM}_1$              & $\mathrm{FOM}_2$      \\ \hline
\begin{tabular}[c]{@{}c@{}}LN WGM resonator \cite{Furst10}\end{tabular} 	& $1.064 - 0.532$                & \multicolumn{2}{c|}{($3.4 \times 10^7$, - )}  & \multicolumn{2}{c|}{($6.8 \times 10^7$, - )}            & -             & $\sim 10^{10}$            & -                 \\ \hline
AlN microring~\cite{Pernice12}                                                                         	& $1.55 - 0.775$                 & \multicolumn{2}{c|}{($\sim 10^4, \sim 5000$)} & \multicolumn{2}{c|}{-}                                  & -             & $2.6 \times 10^5$         & -                 \\ \hline
\begin{tabular}[c]{@{}c@{}}GaP PhC slab~\cite{Rivoire09} $^*$\end{tabular}       		& $1.485 - 0.742$                & \multicolumn{2}{c|}{($\approx 6000$, - )}     & \multicolumn{2}{c|}{-}                                  & -             & $\approx 2 \times 10^5$ & -                 \\ \hline
\multirow{2}{*}{\begin{tabular}[c]{@{}c@{}}GaAs PhC nanobeam~\cite{Buckley14} \end{tabular}} 	& $1.7 - 0.91^\dagger$           & \multicolumn{2}{c|}{(5000, 1000)}             & \multicolumn{2}{c|}{($>10^6$, 4000)}                    & 0.00021       & 820                       & $1.8\times 10^8$  \\ \cline{2-9} 
                                                                                         		& $1.8 - 0.91$                   & \multicolumn{2}{c|}{(5000, 1000)}             & \multicolumn{2}{c|}{($6 \times 10^4$, 4000)}            & 0.00012       & 227                       & $2.1 \times 10^5$ \\ \hline
\begin{tabular}[c]{@{}c@{}}AlGaAs nanoring~\cite{Bi12}\end{tabular}      		& $1.55 - 0.775$                 & \multicolumn{2}{c|}{(5000, 1000)}             & \multicolumn{2}{c|}{($10^4$, $>10^6$)}                  & 0.004         & $10^5$                    & $1.6 \times 10^9$ \\ \hline
\end{tabular}
\caption{SHG figures of merit, including the frequencies $\lambda$, overall and radiative quality factors $Q,Q^\text{rad}$ and nonlinear coupling $\bar{\beta}$ of the fundamental and harmonic modes, of representative geometries. Also shown are the FOM$_1$ and FOM$_2$ figures of merit described in \eqreftwo{F1}{F2}. \newline
  $^*$ SHG occurs between a localized defect mode (at the fundamental frequency) andh an extended index-guided mode of the PhC. \newline $^\dagger$ Resonant frequencies are mismatched. }
\end{table*}

{\it Optimization formulation.---} Optimization techniques have been
regularly employed by the photonic device community, primarily for
fine-tuning the characteristics of a pre-determined geometry; the
majority of these techniques involve probabilistic Monte-Carlo
algorithms such as particle swarms, simulated annealing and genetic
algorithms \cite{Kim04, Behnam10, Minkov14}. While some of these
\emph{gradient-free} methods have been used to uncover a few
unexpected results out of a limited number of degrees of freedom
(DOF)~\cite{Gondarenko06}, \emph{gradient-based} topology optimization
methods efficiently handle a far larger design space, typically
considering every pixel or voxel as a DOF in an extensive 2D or 3D
computational domain, giving rise to novel topologies and geometries
that might have been difficult to conceive from conventional intuition
alone. The early applications of topology optimization were primarily
focused on mechanical problems\cite{Bendose04} and only recently have
they been expanded to consider photonic systems, though largely
limited to \emph{linear} device designs \cite{Gondarenko06, Jensen11,
  Liang13, Liu13, Piggott14, MenLee14}.  In what follows, we describe a
technique for gradient-based topology optimization of nonlinear
wavelength-scale frequency converters.

Recent work~\cite{Liang13} considered topology optimization of the
cavity Purcell factor by exploiting the concept of local density of
states (LDOS). In particular, this previous formulation exploited the
equivalency between LDOS and the power radiated by a \emph{point}
dipole in order to reduce Purcell-factor maximization problems to a
series of small scattering calculations. Defining the objective
$\mathrm{max}_{\bar{\epsilon}}~f\lp \bar{\epsilon}(\vr);\w \rp = -
\operatorname{Re}[\int d\vr \mathbf{J}^* \cdot \vE]$, it follows that
$\vE$ can be found by solving the frequency domain Maxwell's equations
${\cal M} \vE = i \w \mathbf{J}$, where ${\cal M}$ is the Maxwell
operator [SM] and $\mathbf{J}=\delta \lp \vr - \vr_0 \rp
\mathbf{\hat{e}}_j$. The maximization is then performed over a finely
discretized space defined by the \emph{normalized} dielectric
function,~$\{ \bar{\epsilon}_\alpha =
\bar{\epsilon}(\vr_\alpha),~\alpha \leftrightarrow (i \Delta x,j
\Delta y,k \Delta z) \}$. A key realization in~\cite{Liang13} is that
instead of maximizing the LDOS at a single discrete frequency $\w$, a
better-posed problem is that of maximizing the frequency-averaged $f$
in the vicinity of $\w$, denoted by $\langle f \rangle = \int
d\w'~{\cal W}(\w';\w,\Gamma) f(\w')$, where ${\cal W}$ is a weight
function defined over some specified bandwidth $\Gamma$. Using contour
integration techniques, the frequency integral can be conveniently
replaced by a single evaluation of $f$ at a complex frequency $\w + i
\Gamma$~\cite{Liang13}. For a fixed $\Gamma$, the frequency average
effectively shifts the algorithm in favor of minimizing $V$ over
maximizing $Q$; the latter can be enhanced over the course of the
optimization by gradually winding down the averaging bandwidth
$\Gamma$~\cite{Liang13}. A major merit of the frequency-averaged LDOS
formulation is that it features a mathematically well-posed objective
as opposed to a direct maximization of the cavity Purcell factor ${ Q
  \over V}$, allowing for rapid convergence of the optimization
algorithm into an extremal solution.

As suggested in~\cite{Liang13}, a simple extension of the optimization
problem from single- to multi-mode cavities maximizes the minimum of a
collection of LDOS at different frequencies. Applying such an approach
to the problem of SHG, the optimization objective becomes:
$\mathrm{max}_{\bar{\epsilon}_\a} \mathrm{min}
\Big[\mathrm{LDOS}(\w_1),\mathrm{LDOS}(2\w_1)\Big]$, which would
require solving two \emph{separate} scattering problems, ${\cal
  M}_1\vE_1=\mathbf{J}_1$ and ${\cal M}_2\vE_2=\mathbf{J}_2$, for the
two distinct \emph{point} sources $\mathbf{J}_1$, $\mathbf{J}_2$ at
$\w_1$ and $\w_2 = 2 \w_1$ respectively. However, as discussed before,
rather than maximizing the Purcell factor at individual resonances,
the key to realizing optimal SHG is to maximize the overlap integral
$\bar{\beta}$ between $\vE_1$ and $\vE_2$, described by
\eqref{beta}. Here, we suggest an elegant way to incorporate
$\bar{\beta}$ by \emph{coupling} the two scattering problems.  In
particular, we consider not a point dipole but an extended source
$\mathbf{J}_2 \sim \vec{E}_1^2$ at $\w_2$ and optimize a single
\emph{combined} radiated power $f = -\operatorname{Re}\Big[ \int
  d\vr~\mathbf{J}_2^* \cdot \vE_2 \Big]$ instead of two otherwise
\emph{unrelated} LDOS. The advantage of this approach is that $f$
yields precisely the $\bar{\beta}$ parameter along with any resonant
enhancement factors $(\sim Q/V)$ in $\vE_1$ and $\vE_2$. Intuitively,
$\mathbf{J}_2$ can be thought of as a nonlinear polarization current
induced by $\vE_1$ in the presence of the second order susceptibility
tensor $\pmb{\chi}^{(2)}$, and in particular is given by $J_{2i} =
\bar{\epsilon}(\vr) \sum_{jk} \chi^{(2)}_{ijk} E_{1j} E_{1k}$ where
the indices $i,j,k$ run over the Cartesian coordinates. In general,
$\chi^{(2)}_{ijk}$ mixes polarizations and hence $f$ is a sum of
different contributions from various polarization-combinations. In
what follows and for simplicity, we focus on the simplest case in
which $\mathbf{E}_1$ and $\vE_2$ have the same polarization,
corresponding to a diagonal $\bm{\chi}^{(2)}$ tensor determined by a
scalar $\chi^{(2)}_\text{eff}$. Such an arrangement can be obtained
for example by proper alignment of the crystal orientation
axes~[SM]. With this simplification, the generalization of the linear
topology-optimization problem to the case of SHG becomes:
\begin{align}
\text{max}_{\bar{\epsilon}_\a} ~ \langle f(\bar{\epsilon}_\a;\w_1) \rangle &= - \mathrm{Re}\Big[ \Big\langle \int \mathbf{J}_2^* \cdot \mathbf{E}_2 ~d\mathbf{r} \Big \rangle \Big], \label{eq:obj} \\
{\cal M}_1 \mathbf{E}_1 &= i \omega_1 \mathbf{J}_1, \notag \\
{\cal M}_2 \mathbf{E}_2 &= i \omega_2 \mathbf{J}_2,~\w_2 = 2 \w_1 \notag
\end{align}
where
\begin{align}
\mathbf{J}_1 &= \delta(\mathbf{r}_\a-\mathbf{r}_0)
\hat{\mathbf{e}}_j,~j \in \{x,y,z\} \notag \\ \mathbf{J}_2 &=
\bar{\epsilon}(\vr_\a) E_{1j}^2 \hat{\mathbf{e}}_j, \notag \\ {\cal
  M}_l &= \nabla \times {1 \over \mu}~\nabla \times -~
\epsilon_l(\mathbf{r}_\a) \omega_l^2,~l=1,2 \notag
\\ \epsilon_l(\mathbf{r}_\a) &= \epsilon_\text{m} + \bar{\epsilon}_\a
~ \left( \epsilon_{\text{d}l} - \epsilon_\text{m} \right),
~\bar{\epsilon}_\a \in [0,1], \notag
\end{align}
and where $\epsilon_\text{d}$ denotes the dielectric contrast of the
nonlinear medium and $\epsilon_\text{m}$ is that of the surrounding
linear medium. Note that $\bar{\epsilon}_\a$ is allowed to vary
continuously between 0 and 1 whereas the intermediate values can be
penalized by so-called threshold projection filters~\cite{Wang11}. The
scattering framework makes it straightforward to calculate the
derivatives of $f$ (and possible functional constraints) with
respective to $\bar{\epsilon}_\a$ via the adjoint variable
method~\cite{Bendose04, Jensen11,Liang13}. The optimization problem
can then be solved by any of the many powerful algorithms for convex,
conservative, separable approximations, such as the well-known method
of moving asymptotes~\cite{Svanberg02}.

For computational convenience, the optimization is carried out using a
2D computational cell (in the $xz$-plane), though the resulting
optimized structures are given a finite transverse extension $h_y$
(along the $y$ direction) to make realistic 3D devices
(see~\figref{fig3}). In principle, the wider the transverse dimension,
the better the cavity quality factors since they are closer to their
2D limit which only consists of radiation loss in the $z$ direction;
however, as $h_y$ increases, $\bar{\beta}$ decreases due to increasing
mode volumes. In practice, we chose $h_y$ on the order of a few vacuum
wavelengths so as not to greatly compromise either $Q$ or
$\bar{\beta}$. We then analyze the 3D structures via rigorous FDTD
simulations to determine the resonant lifetimes and modal overlaps. By
virtue of our optimization scheme, we invariably find that frequency
matching is satisfied to within the mode linewidths. We note that our
optimization method seeks to maximize the \emph{intrinsic} geometric
parameters such as $Q^\text{rad}$ and $\bar{\beta}$ of an
\emph{un-loaded} cavity whereas the \emph{loaded} cavity lifetime $Q$
depends on the choice of coupling mechanism, e.g. free-space, fiber,
or waveguide coupling, and is therefore an external parameter that can
be considered independently of the optimization. When evaluating the
performance characteristics such as $ \mathrm{FOM}_1 $, we assume
total operational lifetimes $Q_1=5000,~Q_2=1000$. In the optimized
structures, it is interesting to note the appearance of deeply
sub-wavelength features $\sim 1-5\%$ of ${\lambda_1 \over n}$,
creating a kind of \emph{metamaterial} in the optimization direction;
these arise during the optimization process regardless of starting
conditions due to the low-dimensionality of the problem. We find that
these features are not easily removable as their absence greatly
perturbs the quality factors and frequency matching.

{\it Concluding remarks.---} We have presented a formulation that
allows for large-scale optimization of SHG. Applied to simple
micropost and grating structures, our approach yields new classes of
microcavities with stronger performance metrics over existing
designs. One potentially challenging aspect for fabrication in the
case of gratings is the presence of deeply sub-wavelength features,
which would require difficult high-aspect-ratio etching or growth
techniques.  This is not an issue for the micropost cavities since
each material layer can be grown/deposited to a nearly arbitrary
thickness~\cite{Vahala04, Lermer12}.  Another caveat about
wavelength-scale cavities is that they are sensitive to structural
perturbations near the cavity center, where most of the field
resides. In our optimized structures, the figures of merit are robust
to within $\sim \pm 20~\mathrm{nm}$ variations~(approximately one
computational pixel).  One possible way to constrain the optimization
to ensure some minimum spatial feature and robustness is to exploit
so-called regularization filters and worst-case optimization
techniques~\cite{Wang11}, which we will consider in future work.

Our results provide just a glimpse of the kinds of designs that could
be realized in structures with 2D and 3D variations, where we expect
even better performance metrics due to the larger design space. In
fact, preliminary application of our formulation to 2D systems reveals
overlap factors and lifetimes at least one order of magnitude larger
than those attained here. Apart from SHG optimization, our approach
can be generalized to consider other nonlinear processes, even those
involving more than two frequencies [SM]. Preliminary investigations
reveal orders-of-magnitude improvements in the efficiency of third
harmonic and sum-frequency generation processes. These findings,
together with higher-dimensional applications, will be presented in
future work.



\bibliography{opt}

\begin{thebibliography}{10}

\bibitem{DeLong94}
K.~W. DeLong, Rick Trebino, J.~Hunter, and W.~E. White.
\newblock Frequency-resolved optical gating with the use of second-harmonic
  generation.
\newblock {\em J.~Opt. Soc. Am.~B}, 11:2206--2215, 1994.

\bibitem{Arbore97}
M.~A. Arbore, A.~Galvanauskas, D.~Harter, M.~H. Chou, and M.~M. Fejer.
\newblock Engineerable compression of ultrashort pulses by use of
  second-harmonic generation in chirped-period-poled lithium niobate.
\newblock {\em Opt. Lett.}, 22:1341--1343, 1997.

\bibitem{Heinz82}
T.~F. Heinz, C.~K. Chen, D.~Ricard, and Y.~R. Shen.
\newblock Spectroscopy of molecular monolayers by resonant second-harmonic
  generation.
\newblock {\em Phys. Rev. Lett.}, 48:478, 1982.

\bibitem{Kuo06}
P.~S. Kuo, K.~L. Vodopyanov, M.~M. Fejer, D.~M. Simanovskii, X.~Yu, J.~S.
  Harris, D.~Bliss, and D.Weyburne.
\newblock Optical parametric generation of a mid-infrared continuum in
  orientation-patterned gaas.
\newblock {\em Opt. Lett.}, 31:71--73, 2006.

\bibitem{Vodopyanov06}
K.~L. Vodopyanov, M.~M. Fejer, X.~Yu, J.~S. Harris, Y.-S. Lee, W.~C. Hurlbut,
  V.~G. Kozlov, D.~Bliss, and C.~Lynch.
\newblock Terahertz-wave generation in quasi-phase-matched gaas.
\newblock {\em Appl. Phys. Lett.}, 89:141119, 2006.

\bibitem{Krischek10}
Roland Krischek, Witlef Wieczorek, Akira Ozawa, Nikolai Kiesel, Patrick
  Michelberger, Thomas Udem, and Harald Weinfurter.
\newblock Ultraviolet enhancement cavity for ultrafast nonlinear optics and
  high-rate multiphoton entanglement experiments.
\newblock {\em Nature Photonics}, 4:170--173, 2010.

\bibitem{Vaziri02}
Alipasha Vaziri, Gregor Weihs, and Anton Zeilinger.
\newblock Experimental two-photon, three-dimensional entanglement for quantum
  communication.
\newblock {\em Phys. Rev. Lett.}, 89:240401, Nov 2002.

\bibitem{Tanzilli05}
S.~Tanzilli, W.~Tittel, M.~Halder, O.~Alibart, P.~Baldi, N.~Gisin, and
  H.~Zbinden.
\newblock A photonic quantum information interface.
\newblock {\em Nature}, 437:116--120, 2005.

\bibitem{Zaske12}
Sebastian Zaske, Andreas Lenhard, Christian~A. Ke\ss{}ler, Jan Kettler,
  Christian Hepp, Carsten Arend, Roland Albrecht, Wolfgang-Michael Schulz,
  Michael Jetter, Peter Michler, and Christoph Becher.
\newblock Visible-to-telecom quantum frequency conversion of light from a
  single quantum emitter.
\newblock {\em Phys. Rev. Lett.}, 109:147404, Oct 2012.

\bibitem{JoannopoulosJo08-book}
John~D. Joannopoulos, Steven~G. Johnson, Joshua~N. Winn, and Robert~D. Meade.
\newblock {\em Photonic Crystals: Molding the Flow of Light}.
\newblock Princeton University Press, second edition, February 2008.

\bibitem{Soljacic02:bistable}
Marin Solja{\v{c}}i{\'{c}}, Mihai Ibanescu, Steven~G. Johnson, Yoel Fink, and
  J.~D. Joannopoulos.
\newblock Optimal bistable switching in non-linear photonic crystals.
\newblock {\em Phys. Rev. E Rapid Commun.}, 66:055601({R}), 2002.

\bibitem{Soljacic03:OL}
Marin Soljacic, C.~Luo, J.~D. Joannopoulos, and S.~Fan.
\newblock Nonlinear photonic crystal microdevices for optical integration.
\newblock {\em Opt. Lett.}, 28:637--639, 2003.

\bibitem{Yanik03}
M.~F. Yanik, S.~Fan, and M.~Soljacic.
\newblock High-contrast all-optical bistable switching in photonic crystal
  microcavities.
\newblock {\em Appl. Phys. Lett.}, 83:2739--2741, 2003.

\bibitem{Yanik04}
Mehmet~F. Yanik, Shanhui Fan, Marin Solja{\v{c}}i{\'{c}}, , J.~D. Joannopoulos,
  and Yanik.
\newblock All-optical transistor action with bistable switching in a photonic
  crystal cross-waveguide geometry.
\newblock {\em Opt. Lett.}, 68:2506, 2004.

\bibitem{Bravo-AbadRo10}
J.~Bravo-Abad, A.~W. Rodriguez, J.~D. Joannopoulos, P.~T. Rakich, S.~G.
  Johnson, and M.~Soljacic.
\newblock Efficient low-power terahertz generation via on-chip triply-resonant
  nonlinear frequency mixing.
\newblock {\em Appl. Phys. Lett.}, 96:101110, 2010.

\bibitem{Rivoire09}
Kelley Rivoire, Ziliang Lin, Fariba Hatami, W.~Ted Masselink, and Jelena
  Vu\v{c}kovi\'{c}.
\newblock Second harmonic generation in gallium phosphide photonic crystal
  nanocavities with ultralow continuous wave pump power.
\newblock {\em Opt. Express}, 17(25):22609--22615, Dec 2009.

\bibitem{Pernice12}
W.~H.~P. Pernice, C.~Xiong, C.~Schuck, and H.~X. Tang.
\newblock Second harmonic generation in phase matched aluminum nitride
  waveguides and micro-ring resonators.
\newblock {\em Applied Physics Letters}, 100(22), 2012.

\bibitem{Bi12}
Zhuan-Fang Bi, Alejandro~W. Rodriguez, Hila Hashemi, David Duchesne, Marko
  Loncar, Ke-Ming Wang, and Steven~G. Johnson.
\newblock High-efficiency second-harmonic generation in doubly-resonant
  $\chi$(2) microring resonators.
\newblock {\em Opt. Express}, 20(7):7526--7543, Mar 2012.

\bibitem{Buckley14}
Sonia Buckley, Marina Radulaski, Jingyuan~Linda Zhang, Jan Petykiewicz, Klaus
  Biermann, and Jelena Vu\v{c}kovi\'{c}.
\newblock Multimode nanobeam cavities for nonlinear optics: high quality
  resonances separated by an octave.
\newblock {\em Opt. Express}, 22(22):26498--26509, Nov 2014.

\bibitem{Rodriguez07:OE}
Alejandro Rodriguez, Marin Solja{\v{c}}i{\'{c}}, J.~D. Joannopulos, and
  Steven~G. Johnson.
\newblock $\chi^{(2)}$ and $\chi^{(3)}$ harmonic generation at a critical power
  in inhomogeneous doubly resonant cavities.
\newblock {\em Opt. Express}, 15(12):7303--7318, 2007.

\bibitem{Almeida04}
Vilson~R. Almeida, Carlos~A. Barrios, Roberto~R. Panepucci, and Michal Lipson.
\newblock All-optical control of light on a silicon chip.
\newblock {\em Nature}, 431:1081--1084, 2004.

\bibitem{Xu05}
Qianfan Xu and Michal Lipson.
\newblock Carrier-induced optical bistability in silicon ring resonators.
\newblock {\em Opt. Lett.}, 31(3):341--343, 2005.

\bibitem{Levy11}
Jacob~S. Levy, Mark~A. Foster, Alexander~L. Gaeta, and Michal Lipson.
\newblock Harmonic generation in silicon nitride ring resonators.
\newblock {\em Opt. Express}, 19(12):11415--11421, 2011.

\bibitem{Deotare09}
Parag~B. Deotare, Murray~W. McCutcheon, Ian~W. Frank, Mughees Khan, and Marko
  Loncar.
\newblock High quality factor photonic crystal nanobeam cavities.
\newblock {\em Appl. Phys. Lett.}, 94:121106, 2009.

\bibitem{Vahala04}
Kerry Vahala, editor.
\newblock {\em Optical microcavities}.
\newblock World Scientific, USA, 2004.

\bibitem{Lermer12}
M.~Lermer, N.~Gregersen, F.~Dunzer, S.~Reitzenstein, S.~Hofling, J.~Mork,
  L.~Worschech, M.~Kamp, and A.~Forchel.
\newblock Bloch-wave engineering of quantum dot micropillars for cavity quantum
  electrodynamics experiments.
\newblock {\em Phys. Rev. Lett.}, 108:057402, 2012.

\bibitem{YZhang09}
Yinan Zhang and Marko Loncar.
\newblock Sub-micron diameter micropillar cavities with high quality factors
  and ultra-small mode volumes.
\newblock {\em arXiv:0902.2553}, 2009.

\bibitem{Scalora97}
M.~Scalora, M.~J. Bloemer, A.~S. Manka, J.~P. Dowling, C.~M. Bowden,
  R.~Viswanathan, and J.~W. Haus.
\newblock Pulsed second-harmonic generation in nonlinear, one-dimensional,
  periodic structures.
\newblock {\em Phys. Rev.~A}, 56:3166, 1997.

\bibitem{Furst10}
J.~U. F\"urst, D.~V. Strekalov, D.~Elser, M.~Lassen, U.~L. Andersen,
  C.~Marquardt, and G.~Leuchs.
\newblock Naturally phase-matched second-harmonic generation in a
  whispering-gallery-mode resonator.
\newblock {\em Phys. Rev. Lett.}, 104:153901, Apr 2010.

\bibitem{Diziain13}
Severine Diziain, Reinhard Geiss, Matthias Zilk, Frank Schrempel,
  Ernst-Bernhard Kley, Andreas Tunnermann, and Thomas Pertsch.
\newblock Second harmonic generation in free-standing lithium niobate photonic
  crystal l3 cavity.
\newblock {\em Applied Physics Letters}, 103(5):051117--051117--4, Jul 2013.

\bibitem{Wang14}
Cheng Wang, Michael~J. Burek, Zin Lin, Haig~A. Atikian, Vivek Venkataraman,
  I-Chun Huang, Peter Stark, and Marko Lon\v{c}ar.
\newblock Integrated high quality factor lithium niobate microdisk resonators.
\newblock {\em Opt. Express}, 22(25):30924--30933, Dec 2014.

\bibitem{Kuo14}
Kuo~P. S., Bravo-Abad J., and Solomon~G. S.
\newblock Second-harmonic generation using quasi-phasematching in a gaas
  whisphering-gallery-mode microcavity.
\newblock {\em Nature Comm.}, 5(3109), 2014.

\bibitem{Boyd92}
Robert~W. Boyd.
\newblock {\em Nonlinear Optics}.
\newblock Academic Press, California, 1992.

\bibitem{Miller97}
G.~D. Miller, R.~G. Batchko, W.~M. Tulloch, D.~R. Weise, M.~M. Fejer, and R.~L.
  Byer.
\newblock 42\%-efficient single-pass cw second-harmonic generation in
  periodically poled lithium niobate.
\newblock {\em Opt. Lett.}, 22(24):1834--1836, 1997.

\bibitem{Kim04}
Woo~Jun Kim and John~D. O'Brien.
\newblock Optimization of a two-dimensional photonic-crystal waveguide branch
  by simulated annealing and the finite-element method.
\newblock {\em J. Opt. Soc. Am. B}, 21(2):289--295, Feb 2004.

\bibitem{Behnam10}
Behnam~Saghirzadeh Darki and Nosrat Granpayeh.
\newblock Improving the performance of a photonic crystal ring-resonator-based
  channel drop filter using particle swarm optimization method.
\newblock {\em Optics Communications}, 283(20):4099 -- 4103, 2010.

\bibitem{Minkov14}
Momchil Minkov and Vincenzo Savona.
\newblock Automated optimization of photonic crystal slab cavities.
\newblock {\em Sci. Rep.}, 4(10.1038/srep05124), 2014.

\bibitem{Gondarenko06}
Alexander Gondarenko, Stefan Preble, Jacob Robinson, Long Chen, Hod Lipson, and
  Michal Lipson.
\newblock Spontaneous emergence of periodic patterns in a biologically inspired
  simulation of photonic structures.
\newblock {\em Phys. Rev. Lett.}, 96:143904, Apr 2006.

\bibitem{Bendose04}
Martin~Philip Bendose and Ole Sigmund.
\newblock {\em Topology optimization}.
\newblock Springer, USA, 2004.

\bibitem{Jensen11}
J.S. Jensen and O.~Sigmund.
\newblock Topology optimization for nano-photonics.
\newblock {\em Laser and Photonics Reviews}, 5(2):308--321, 2011.

\bibitem{Liang13}
Xiangdong Liang and Steven~G. Johnson.
\newblock Formulation for scalable optimization of microcavities via the
  frequency-averaged local density of states.
\newblock {\em Opt. Express}, 21(25):30812--30841, Dec 2013.

\bibitem{Liu13}
David Liu, Lucas~H. Gabrielli, Michal Lipson, and Steven~G. Johnson.
\newblock Transformation inverse design.
\newblock {\em Opt. Express}, 21(12):14223--14243, Jun 2013.

\bibitem{Piggott14}
Alexander~Y. Piggott, Jesse Lu, Thomas~M. Babinec, Konstantinos~G. Lagoudakis,
  Jan Petykiewicz, and Jelena Vuckovic.
\newblock Inverse design and implementation of a wavelength demultiplexing
  grating coupler.
\newblock {\em Sci. Rep.}, 4(10.1038/srep07210), 2014.

\bibitem{MenLee14}
Han Men, Karen Y.~K. Lee, Robert~M. Freund, Jaime Peraire, and Steven~G.
  Johnson.
\newblock Robust topology optimization of three-dimensional photonic-crystal
  band-gap structures.
\newblock {\em Optics Express}, 22:22632--22648, September 2014.

\bibitem{Wang11}
Fengwen Wang, BoyanStefanov Lazarov, and Ole Sigmund.
\newblock On projection methods, convergence and robust formulations in
  topology optimization.
\newblock {\em Structural and Multidisciplinary Optimization}, 43(6):767--784,
  2011.

\bibitem{Svanberg02}
Krister Svanberg.
\newblock A class of globally convergent optimization methods based on
  conservative convex separable approximations.
\newblock {\em SIAM Journal on Optimization}, pages 555--573, 2002.

\end{thebibliography}


\begin{thebibliography}{1}

\bibitem{Rodriguez07:OE}
Alejandro Rodriguez, Marin Solja{\v{c}}i{\'{c}}, J.~D. Joannopulos, and
  Steven~G. Johnson.
\newblock $\chi^{(2)}$ and $\chi^{(3)}$ harmonic generation at a critical power
  in inhomogeneous doubly resonant cavities.
\newblock {\em Opt. Express}, 15(12):7303--7318, 2007.

\bibitem{Boyd92}
Robert~W. Boyd.
\newblock {\em Nonlinear Optics}.
\newblock Academic Press, California, 1992.

\bibitem{Liang13}
Xiangdong Liang and Steven~G. Johnson.
\newblock Formulation for scalable optimization of microcavities via the
  frequency-averaged local density of states.
\newblock {\em Opt. Express}, 21(25):30812--30841, Dec 2013.

\bibitem{Bendose04}
Martin~Philip Bendose and Ole Sigmund.
\newblock {\em Topology optimization}.
\newblock Springer, USA, 2004.

\bibitem{Jensen11}
J.S. Jensen and O.~Sigmund.
\newblock Topology optimization for nano-photonics.
\newblock {\em Laser and Photonics Reviews}, 5(2):308--321, 2011.

\bibitem{Bi12}
Zhuan-Fang Bi, Alejandro~W. Rodriguez, Hila Hashemi, David Duchesne, Marko
  Loncar, Ke-Ming Wang, and Steven~G. Johnson.
\newblock High-efficiency second-harmonic generation in doubly-resonant
  $\chi$(2) microring resonators.
\newblock {\em Opt. Express}, 20(7):7526--7543, Mar 2012.

\end{thebibliography}

\end{document}


\title{Cavity-enhanced second harmonic generation via nonlinear-overlap optimization: Supplementary materials}

\author{Zin Lin$^1$}
\email{zlin@seas.harvard.edu}
\author{Xiangdong Liang$^2$}
\author{Marko Loncar$^1$}
\author{Steven G. Johnson$^2$}
\author{Alejandro W. Rodriguez$^{3}$}%
\affiliation{$^1$School of Engineering and Applied Sciences, Harvard University, Cambridge, MA 02138}
\affiliation{$^2$Department of Mathematics, Massachusetts Institute of Technology, Cambridge, MA 02139}
\affiliation{$^3$Department of Electrical Engineering, Princeton University, Princeton, NJ, 08544}

\date{\today}

\begin{abstract}
  We review the temporal-coupled mode equations describing second
  harmonic generation in doubly resonant cavities and motivate the
  dimensionless nonlinear coupling $\bar{\beta}$ described in Eq.~3 of
  the main text. We provide further details on the topology
  optimization formulation for second harmonic generation and describe
  generalizations to other nonlinear processes. Finally, we present
  more detailed descriptions of the optimized micropost and gratings
  cavities.
\end{abstract}

\maketitle

\section{Coupled-mode theory for second harmonic generation}

The temporal coupled mode equations describing second harmonic
generation (SHG) in a doubly-resonant cavity coupled to a channel
are~\cite{Rodriguez07:OE}:
\begin{align}
  {da_1 \over dt} &= i \w_1 \lp 1 + {i \over 2 Q_1} \rp a_1 - i \w_1
  \beta_1 a_1^* a_2 \notag \\ &+ \sqrt{\w_1 \lp {1 \over Q_1} - { 1
      \over Q_1^\text{rad}} \rp} s_{1+}, \label{eq:eq1}\\ {da_2 \over
    dt} &= i \w_2 \lp 1 + {i \over 2 Q_2} \rp a_1 - i \w_2 \beta_2
  a_1^2, \label{eq:eq2} \\ 
  s_{1-} &= \sqrt{\w_1 \lp {1 \over Q_1} - {1 \over Q_1^\text{rad}} \rp} a_1 - s_{1+}, \\
  s_{2-} &= \sqrt{\w_2 \lp {1 \over Q_2} - {1 \over Q_2^\text{rad}} \rp} a_2
\end{align}
such that $|a_k|^2$ is the modal energy in the cavity and $|s_{k
  \pm}|^2$ is the input/output power in the waveguide, and where
$Q_{k}$ and $Q_k^\text{rad}$ denote the total and radiative quality
factors corresponding to mode $k$. The nonlinear coupling coefficient
$\beta_1$, obtained from perturbation theory~\cite{Rodriguez07:OE}, is
given by:
\begin{align}
\beta_1 &= {1 \over 4} { \int d\vr~ \epsilon_0 \sum_{ijk}
  \chitwo_{ijk}(\vr) \lp E_{1i}^* E_{2j}E_{1k}^* + E_{1i}^* E_{1j}^*
  E_{2k} \rp \over \left(\int d\vr~ \epsilon_0 \epsilon_1(\vr)
  |\mathbf{E}_1|^2\right) \left( \sqrt{\int d\vr~ \epsilon_0
    \epsilon_2(\vr) |\mathbf{E}_2|^2 } \right) }. \notag
\end{align}
with $\beta_2 = \beta_1^*/2$ far off from material resonances where
Kleinman symmetry is valid~\cite{Boyd92}, as required by conservation
of energy~\cite{Rodriguez07:OE}. In general, the overlap integral in
the numerator is a sum of products between different $E$-field
polarizations weighted by off-diagonal components of the nonlinear
$\chi^{(2)}$ tensor. For simplicity, however, in the main text we only
consider the simple case of diagonal $\chi^{(2)}$ involving
same--polarization interactions described by an effective
$\chi^{(2)}_\text{eff}$, resulting from an appropriate orientation of
the crystal axes of the nonlinear material. All of these
considerations suggest a simple dimensionless normalization of
$\beta$, given by:
\begin{align}
  \bar{\beta} = { \int d\vr ~ \bar{\epsilon}(\vr) E_2^* E_1^2 \over
    \left( \int d\vr~ \epsilon_1 |\mathbf{E}_1|^2 \right) \left(
      \sqrt{\int d\vr~ \epsilon_2 |\mathbf{E}_2|^2 } \right) } \sqrt{
    \lambda_1^3},
\end{align}
such that $\beta_2 = 4 \bar{\beta} \chi^{(2)}_\text{eff} /
\sqrt{\epsilon_0 \lambda_1^3}$. As defined in the text,
$\bar{\epsilon}(\vr) = 1$ for nonlinear dielectric and
$\bar{\epsilon}(\vr) = 0$ for the surrounding linear medium.

Most SHG experiments operate in the small-signal regime of small input
powers, leading to negligible down-conversion and pump
depletion. Ignoring the down-conversion or $\beta_1$ term in
\eqref{eq1}, one obtain the following simple expression for the second
harmonic output power:
\begin{align} 
{P_2^\text{out} \over \lp P_1^\text{in} \rp^2 } = { 8
    \over \w_1} \lp { \chi^{(2)}_\text{eff} \over \sqrt{\epsilon_0
      \lambda_1^3}} \rp^2 ~Q_1^2 Q_2 |\bar{\beta}|^2 \lp 1 - { Q_1
    \over Q_1^\text{rad}} \rp^2 \lp 1 - { Q_2 \over Q_2^\text{rad}}
  \rp.
\end{align}
\normalsize In the limit of large up-conversion and non-negligible
down-conversion, solution of the coupled-mode equations yields the
maximum efficiency (defined as $\eta = P_2^\text{out}/P_1^\text{in}$) and corresponding critical
power~\cite{Rodriguez07:OE}:
\begin{align}
  \eta^\text{max} &= \lp 1 - {Q_1 \over Q_1^\text{rad}} \rp \lp 1 - {Q_2 \over Q_2^\text{rad}} \rp, \\
  P_1^\text{crit} &= {2 \w_1 \epsilon_0 \lambda_1^3 \over \lp
    \chi^{(2)}_\text{eff} \rp ^2 } {1 \over |\bar{\beta}|^2 Q_1^2 Q_2}
  \lp 1 - {Q_1 \over Q_1^\text{rad}} \rp^{-1}.
\end{align}

\begin{figure}[t!]
\centering
\includegraphics[width=\columnwidth]{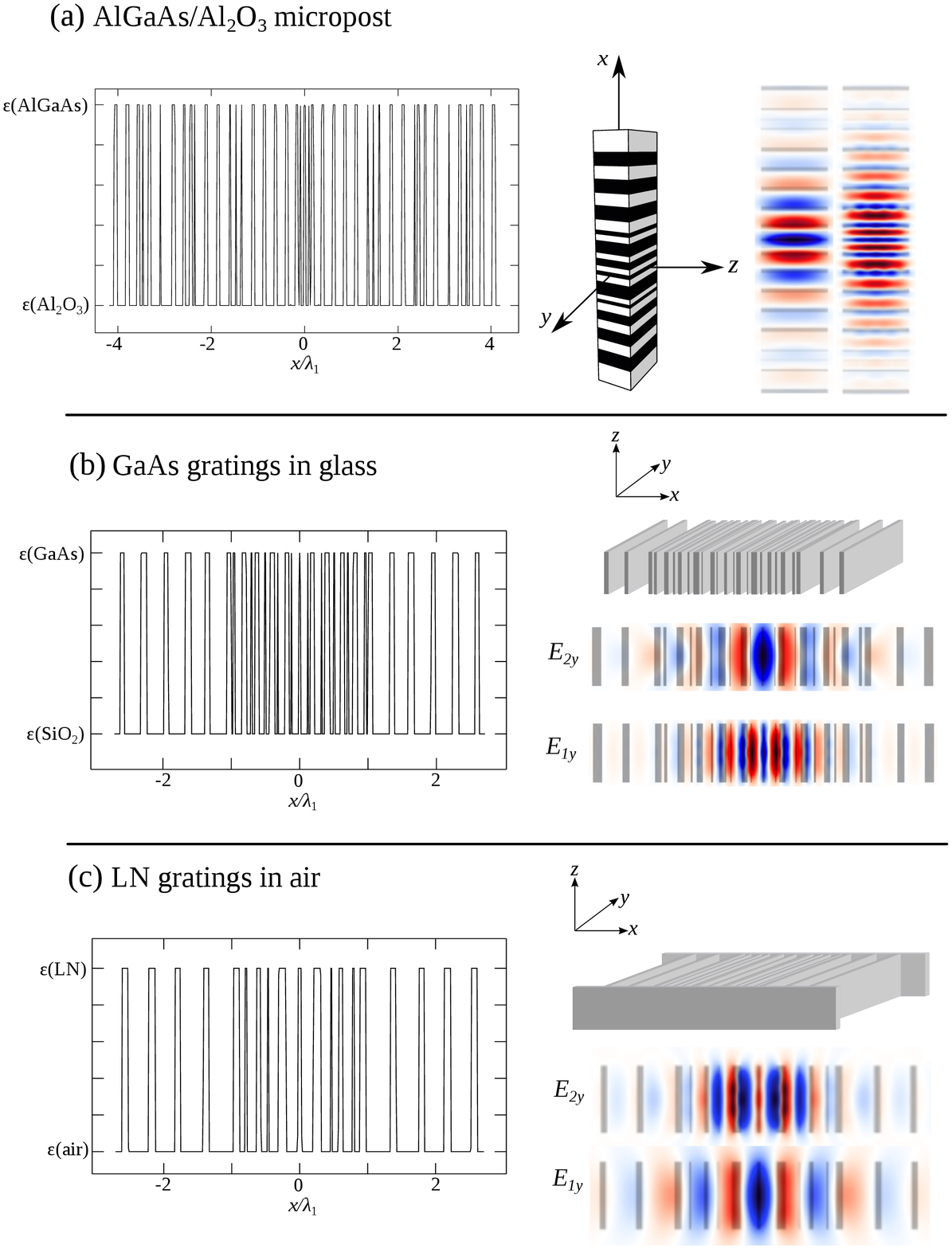}
\caption{Cross-sectional dielectric profiles and electric field
  distributions for AlGaAs/Al$_2$O$_3$ micropost (a), GaAs gratings in
  SiO$_2$ (b) and LN gratings in air (c).}
\label{fig:fig1}
\end{figure}

\section{Formulation for topology optimization of arbitrary nonlinear
  frequency conversion process}

Nonlinear frequency conversion processes can be viewed as frequency
mixing schemes in which two or more \emph{constituent} photons at a
set of frequencies $\{\w_n\}$ interact to produce an output photon at
frequency $\Omega$ such that $ \Omega=\sum_n c_n \w_n$, where the
photon number coefficients $\{c_n\}$ can be either negative or
positive, depending on whether the corresponding photons are created
or destroyed in the process. In general, because the optical nonlinear
response of materials is a tensor and hence the frequency conversion
process mixes different polarizations~\cite{Boyd92}. However, for
notational simplicity, we will describe our optimization problem only
for a single component of the susceptibility tensor. If one wishes to
consider the full tensorial properties, one can easily add extra
optimization terms (weighted by the tensor components) by following
the same approach described below.

Given a specific nonlinear tensor component $\chi_{ijk...}$, where $i,j,k,...\in\{x,y,z\}$,
mediating an interaction between the polarization components $E_i(\Omega)$ and $E_{1j}$, $E_{2k}, ...$, 
we begin with a collection of point dipole currents, each at the
\emph{constituent} frequency $\w_n,~n\in\{1,2,...\}$ and positioned at the center of
the computational cell $\mathbf{r}'$, such that $\mathbf{J}_n =
\hat{\mathbf{e}}_{n \nu} \delta(\mathbf{r}-\mathbf{r}')$, where
$\hat{\mathbf{e}}_{n \nu} \in\{\hat{\mathbf{e}}_{1j},~\hat{\mathbf{e}}_{2k}, ...\}$ is a polarization vector chosen so as to excite
the desired $E$-field polarization components of the corresponding
mode. Given the choice of incident currents $\mathbf{J}_n$, we solve
Maxwell's equations to obtain the corresponding \emph{constituent}
electric-field response $\vec{E}_n$, from which one can construct a
nonlinear polarization current $\mathbf{J}(\Omega) =
\bar{\epsilon}(\mathbf{r}) \prod_{n} E_{n\nu}^{|c_n| (*)}
\hat{\mathbf{e}}_i$, where $E_{n\nu}= \vE_n \cdot \hat{\mathbf{e}}_{n\nu}$
and $\mathbf{J}(\Omega)$ has polarization
$\hat{\mathbf{e}}_i$ generally different from the constituent polarizations $\hat{\mathbf{e}}_{n\nu}$. 
Here, (*) denotes complex conjugation for negative $c_n$ and no conjugation otherwise.
Finally, we maximize the radiated power $- \mathrm{Re}\Big[
  \int \mathbf{J}(\Omega)^* \cdot \mathbf{E}(\Omega) ~d\mathbf{r}
  \Big]$ from $\mathbf{J}(\Omega)$.

The formulation is now given by:
\begin{align}
  \text{max}_{\bar{\epsilon}} ~ f(\bar{\epsilon};\w_n) &= - \mathrm{Re}\Big[ \int \mathbf{J}(\Omega)^* \cdot \mathbf{E}(\Omega) ~d\mathbf{r} \Big], \label{eq:ps1}\\
  {\cal M}(\bar{\epsilon},\omega_n) \mathbf{E}_n &= i \omega_n \mathbf{J}_n,~ \mathbf{J}_n = \hat{\mathbf{e}}_{n \nu} \delta(\mathbf{r}-\mathbf{r}'), \notag \\
  {\cal M}(\bar{\epsilon},\Omega) \mathbf{E}(\Omega) &= i \Omega \mathbf{J}(\Omega),~ \mathbf{J}(\Omega) = \bar{\epsilon} \prod_{n} E_{n \nu}^{|c_n| (*)} \hat{\mathbf{e}}_i, \notag \\
  {\cal M}(\bar{\epsilon},\omega) &= \nabla \times {1 \over \mu}~\nabla \times -~ \epsilon(\mathbf{r}) \omega^2, \notag \\
  \epsilon(\mathbf{r}) &= \epsilon_\text{m} + \bar{\epsilon} ~ \left( \epsilon_\text{d} - \epsilon_\text{m} \right), ~\bar{\epsilon} \in [0,1]. \notag
\end{align} 
In practice, we maximize a frequency-averaged version of the power
output $\langle f(\w) \rangle$ rather than $f(\w)$ itself since the
latter has poor convergence \cite{Liang13}, i.e., we maximize $\langle
f \rangle = \int d\w'~{\cal W}(\w';\w,\Gamma) f(\w')$, where we simply
choose the weighting function ${\cal W}$ to be a simple lorentzian
with the desired resonance $\w$ and a certain linewidth
$\Gamma$,
\begin{align}
{\cal W}(\w') = { \Gamma / \pi \over (\w' - \w )^2 + \Gamma^2 } 
\end{align}
(Note that this linewidth is not necessarily the same as the intrinsic
radiative linewidth of the cavity in that $\Gamma$ is a computational
artifice introduced to aid rapid convergence~\cite{Liang13}.) In
Ref. \cite{Liang13}, it is shown by means of contour integration that
this averaging is equivalent to evaluating $f$ at a complex frequency
$f(\w+i \Gamma)$, also equivalent to adding a uniform loss $i
\Gamma/\w$ to $\mu (\vr)$ and $\epsilon(\vr)$. In our implementation
of the optimization process, we typically begin with a large $\Gamma$
which affords rapid convergence to a stable geometry in a few hundred
iterations. $\Gamma$ is then decreased by an order of magnitude every
time the optimization converges until $\Gamma \sim 10^{-5}$ at which
point the structure settles to within a linewidth $\Gamma$ of the
desired frequencies (perfect frequency matching).


Application of this formulation to the problem of second harmonic
generation is straightforward and described in the main text, in which
case $\Omega = \w_2 = 2 \w_1$ and $\mathbf{J}(\Omega) = \mathbf{J}_2 =
\bar{\epsilon}(\mathbf{r})~(\hat{\mathbf{e}}_{1j}\cdot\mathbf{E}_1)^2
\hat{\mathbf{e}}_{2i}$. Note that for the structures described in the
text, we chose $\hat{\mathbf{e}}_{1j} = \hat{\mathbf{e}}_{2i} =
\hat{\mathbf{e}}_y$. In addition to the problem statement of
\eqref{ps1}, the optimization algorithm also benefits from gradient
information of the objective function, which exploits the adjoint
variable method \cite{Bendose04, Jensen11, Liang13}. Here, we simply
quote the result for the gradient of our SHG objective function (dropping the polarization index $y$ for simplicity),
$\langle f(\bar{\epsilon};\w_1) \rangle = - \mathrm{Re}\Big[
  \Big\langle \int \mathbf{J}_2^* \cdot \mathbf{E}_2 ~d\mathbf{r} \Big
  \rangle \Big]$,
\begin{align}
  {\partial \langle f \rangle \over \partial \bar{\epsilon} } = &-\operatorname{Re} \Big[
  E_2 \lp E_1^* \rp^2 + ( \epsilon_{\text{d}1} - \epsilon_\text{m}) \w_1^2 u_1^* E_1^* \notag \\
  &+ ( \epsilon_{\text{d}2} - \epsilon_\text{m})\w_2^2 u_2 E_2 
  + i \w_2 u_2 E_1^2 \notag \\
  &+ i \w_2 \w_1^2 (\epsilon_{\text{d}1} - \epsilon_\text{m}) u_3 E_1 \Big],  \notag
\end{align}
\normalsize where the functions $u_k$ are solutions of the following
scattering problems:
\begin{align}
{\cal M}_1 u_1 &= \bar{\epsilon} E_1^* E_2, \notag \\
{\cal M}_2 u_2 &= \bar{\epsilon} E_1^2, \notag \\
{\cal M}_1 u_3 &= 2 \bar{\epsilon} E_1^* u_2. \notag
\end{align}

\section{Optimized 1D cavity designs}

\Figref{fig1} shows the dielectric and $E_y$ field profiles of the
three optimized structures, including (a) AlGaAs/Al$_2$O$_3$
micropost, (b) GaAs gratings in SiO$_2$, and (3) LN gratings in
air. Along the $x$ cross-section, each computational pixel of
thickness $\Delta$ represents \emph{either} a high dielectric (nonlinear) or
low dielectric (linear) material. For example, in the
AlGaAs/Al$_2$O$_3$ micropost cavity (assuming 
$n_1(\text{AlGaAs})=3.02$ and $n_2(\text{AlGaAs})=3.18$ for AlGaAs with 70\% Al~\cite{Bi12}, and 
$n(\text{Al$_2$O$_3$})=1.7$), we took $\Delta = 0.015~\lambda_1$.

\bibliography{opt}